\documentclass[sigconf]{acmart}

\makeatletter
\renewcommand\@formatdoi[1]{\ignorespaces}
\makeatother

\usepackage{booktabs} 
\usepackage{braket}
\usepackage{amsmath}
\usepackage{amssymb}
\usepackage{bbm}
\usepackage{subcaption}
\usepackage{pbox}
\usepackage{multirow}

\graphicspath{{figures/plots/}{figures/}{figures/gates/}}

\makeatletter
\def\hlinewd#1{%
\noalign{\ifnum0=`}\fi\hrule \@height #1 %
\futurelet\reserved@a\@xhline}
\makeatother

\begin{document}

\title{0.5 Petabyte Simulation of a 45-Qubit Quantum Circuit}


\author{Thomas H\"aner}
\affiliation{\institution{Institute for Theoretical Physics}\streetaddress{ETH Zurich}\city{8093 Zurich, Switzerland}}
\email{haenert@phys.ethz.ch}
\author{Damian S. Steiger}
\affiliation{\institution{Institute for Theoretical Physics}\streetaddress{ETH Zurich}\city{8093 Zurich, Switzerland}}
\email{dsteiger@phys.ethz.ch}

\begin{abstract}
Near-term quantum computers will soon reach sizes that are challenging to directly simulate, even when employing the most powerful supercomputers. Yet, the ability to simulate these early devices using classical computers is crucial for calibration, validation, and benchmarking. In order to make use of the full potential of systems featuring multi- and many-core processors, we use automatic code generation and optimization of compute kernels, which also enables performance portability. We apply a scheduling algorithm to quantum supremacy circuits in order to reduce the required communication and simulate a $45$-qubit circuit on the Cori II supercomputer using $8,192$ nodes and $0.5$ petabytes of memory. To our knowledge, this constitutes the largest quantum circuit simulation to this date. Our highly-tuned kernels in combination with the reduced communication requirements allow an improvement in time-to-solution over state-of-the-art simulations by more than an order of magnitude at every scale.
\end{abstract}

%
%

\ccsdesc[500]{Applied computing~Physics}

\settopmatter{printacmref=false} 
\renewcommand\footnotetextcopyrightpermission[1]{} 
\pagestyle{plain} 

\maketitle

\section{Introduction}
Quantum computers promise to solve problems which would be impossible to tackle with classical machines. While such devices will not speed up every application, there are certain areas which could be revolutionized by quantum computers. This includes quantum chemistry~\cite{aspuru2005simulated,babbush2015chemical,reiher2016}, material science~\cite{bauer2015hybrid}, machine learning~\cite{harrow2009,wiebe2012,lloyd2013,rebentrost2014}, and cryptography~\cite{shor94}. 

\begin{figure*}[ht]
	\includegraphics[width=\linewidth]{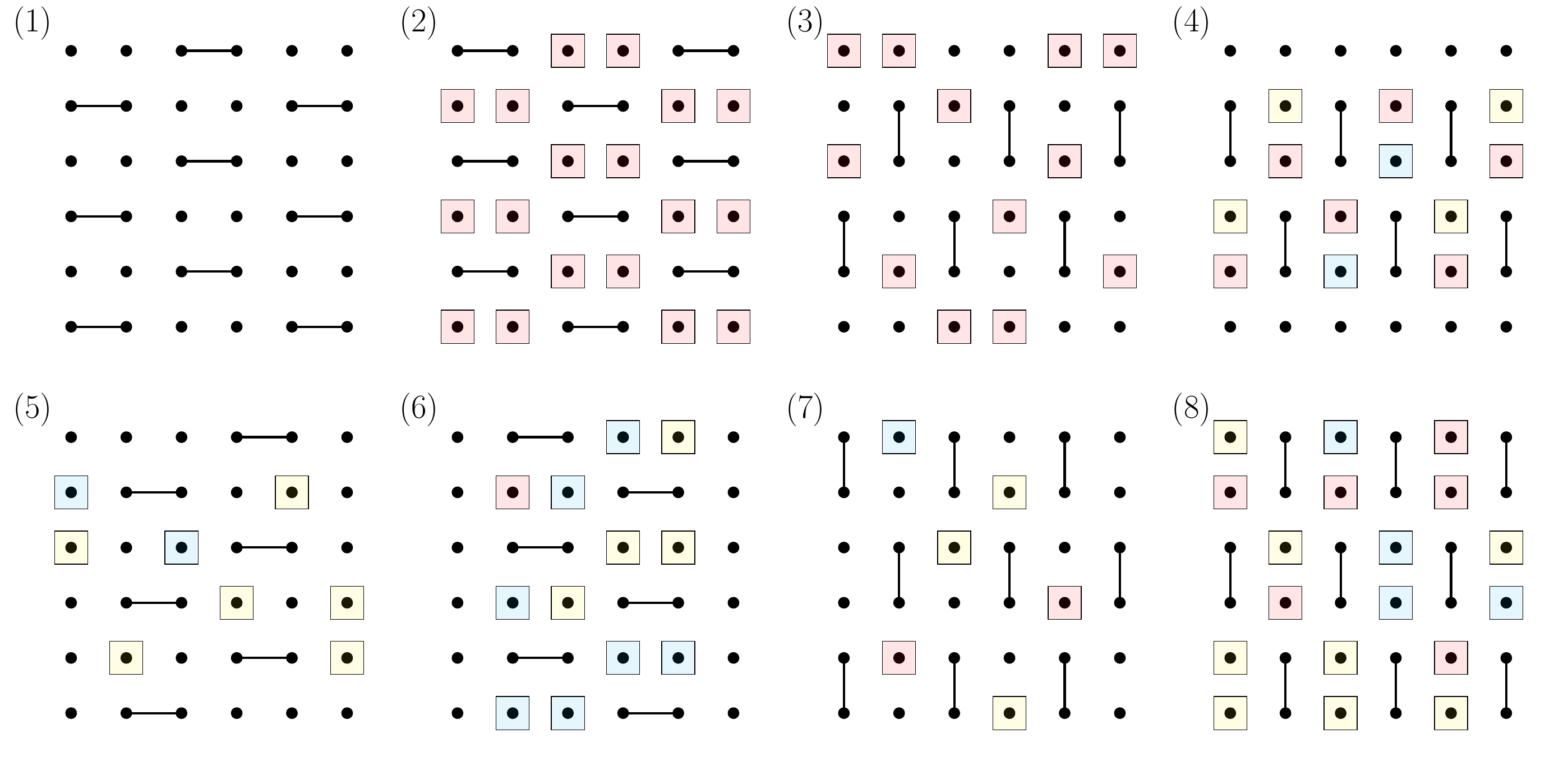}
	\caption{Low-depth random quantum circuit proposed by Google to show quantum supremacy \cite{boixo2016characterizing}. We generated identical circuits using the following rules: At clock cycle 0, a Hadamard gate is applied to each qubit. Afterwards, eight different patterns of controlled Z (CZ) gates are applied repeatedly until the desired circuit depth is achieved. See the 8 different CZ patterns above in clock cycles 1-8 for a $6\times6$ qubit circuit, where the CZ gates are represented by a line between two qubits. This pattern ensures that all possible two qubit interactions on this 2D nearest neighbor architecture are executed every 8 cycles.
		In addition to the CZ gates, single qubit gates are applied to all qubits which in the previous cycle (but not in the current cycle) performed a CZ gate. The single qubit gates are randomly chosen to be either a $T$ (red), $X^{1/2}$ (blue), or $Y^{1/2}$ (yellow) gate, except that the second single-qubit gate on each qubit (the first is the Hadamard gate in cycle 0) is always a T gate and when randomly choosing a single-qubit gate, it must be different from the previous single-qubit gate on that qubit.}
	\label{fig:supremacy_circuit}
\end{figure*}

Experimental devices featuring close to 50 quantum bits (qubits) will soon be available and may be able to perform well-defined computational tasks which would classically require the world's most powerful supercomputers. Going even beyond these capabilities means entering the domain of \textit{Quantum Supremacy}~\cite{boixo2016characterizing,aaronson2016}. While one of the computational tasks proposed to demonstrate this supremacy -- the execution of low-depth random quantum circuits, see Fig.~\ref{fig:supremacy_circuit} -- is not scientifically useful on its own, running such circuits is still of great use to calibrate, validate, and benchmark near-term quantum devices~\cite{boixo2016characterizing}.

Therefore, in addition to verifying quantum algorithms and carrying out studies of their behavior under noise, quantum circuit simulators may provide the means to carry out these calibrations and benchmarks and thereby enable more efficient quantum hardware/software co-design. Quantum circuit simulators are thus comparable to tools such as the structural simulation toolkit (SST)~\cite{rodrigues2011structural}, which allows to simulate upcoming classical hardware.

\paragraph{Related work.} There are many implementations of quantum circuit simulators~\cite{qclist} available, most of them are meant to simulate small systems on a single node. The most widely-used single-node simulator is Microsoft's LIQ$Ui\Ket{}$~\cite{wecker14}, which is implemented in F$\#$ and is thus not as fast as simulators written in, e.g., C++ such as ProjectQ~\cite{steiger2016projectq}. The massively parallel simulator from~\cite{trieu2009large,de2007massively} was used to simulate $42$ qubits on the J\"ulich supercomputer in 2010, which set the new world record in number of simulated quantum bits. Recently, Intel's qHiPSTER~\cite{smelyanskiy2016qhipster} was specialized for the simulation of quantum supremacy circuits and then used to simulate these circuits up to 42 qubits~\cite{boixo2016characterizing}.
A topic similar to quantum circuit simulation is emulation which, instead of simulating individual gates, uses classical shortcuts in order to reduce the time-to-solution for quantum operations whose action is known in advance. An example for such a shortcut is the quantum Fourier transform, which can be emulated by applying a fast Fourier transform to the state vector~\cite{haner2016high}. However, such emulation techniques are not applicable to quantum supremacy circuits.

\paragraph{Our contribution.} We improve the strong scaling behavior of the compute kernels underlying quantum circuit simulation in order to reduce time-to-solution when employing multi- and many-core processors. In the multi-node domain, we employ a communication scheme similar to~\cite{de2007massively} and introduce an additional layer of optimization to reduce the amount of communication required: We apply a clustering algorithm to the quantum circuit in order to improve the scheduling of quantum gate operations. While this pre-computation terminates in $1-3$ seconds on a laptop, it greatly reduces the number of communication steps. We then simulate quantum supremacy circuits of various sizes and report speedups of over one order of magnitude on all scales. Finally, we simulate a 45-qubit quantum supremacy circuit on the Cori II supercomputing system using 0.5 petabytes of memory and $8,192$ nodes. To our knowledge, this constitutes a new record in the maximal number of simulated qubits.
The classical simulation of such circuits is believed to be impossible already for 49 qubits which, according to~\cite{Mohseni2017}, is the threshold for quantum computers outperforming the largest supercomputers available today at the task of sampling from the output distribution of random low-depth quantum circuits. While we do not carry out a classical simulation of 49 qubits, we provide numerical evidence that this may be possible. Our optimizations allow reducing the number of communication steps required to simulate the entire circuit to just two all-to-alls, making it possible to use, e.g., solid-state drives if the available memory is less than the 8 petabytes required.

\section{Basics of quantum computer simulation}
A quantum computer consists of quantum bits (qubits). A qubit is a two-level quantum systems whose state $\Ket{\psi}$ can be described by 2 complex numbers $\alpha_0$ and $\alpha_1$ as
\[
	\Ket{\psi}=\alpha_0\Ket 0 + \alpha_1\Ket 1\;,
\]
where $|\alpha_0|^2+|\alpha_1|^2=1$. The two computational basis states $\Ket 0$ and $\Ket 1$ are orthonormal and the qubit, when measured, collapses onto either $\Ket 0$ or $\Ket 1$ with probability $p=|\alpha_0|^2$ and $|\alpha_1|^2=1-p$, respectively. The measured qubit is then just a classical bit. To store $\Ket{\psi}$ on a classical computer, it is more practical to choose two orthonormal vectors to represent $\Ket 0$ and $\Ket 1$,
\[
	\Ket{\psi}=\alpha_0\left(\begin{matrix}
	1\\0
	\end{matrix}\right)+\alpha_1\left(\begin{matrix}
	0\\1
	\end{matrix}\right)
	=\left(\begin{matrix}
	\alpha_0\\\alpha_1
	\end{matrix}\right)\;.
\]
Any operation on this qubit can then be represented as a complex unitary $2\times 2$ matrix. Applying, for example, a bit-flip operation, denoted by $X=\left(
\begin{smallmatrix}
0 & 1 \\ 1 & 0
\end{smallmatrix}\right)$, to the state $\Ket{\psi}$, amounts to a matrix-vector multiplication:
\[
	X\Ket{\psi}=\left(\begin{matrix}
	0 & 1\\1 & 0
	\end{matrix}\right)\left(\begin{matrix}
	\alpha_0\\\alpha_1
	\end{matrix}\right)
	=
	\left(\begin{matrix}
	\alpha_1\\\alpha_0
	\end{matrix}\right)
	=
	\alpha_1\Ket 0 + \alpha_0\Ket 1\;.
\]
Other single-qubit gates used in quantum supremacy circuits are the Hadamard gate $H=\frac 1{\sqrt 2}\left(\begin{smallmatrix}
	1 & 1 \\ 1 & -1
	\end{smallmatrix}\right)$, the T gate $T=\left(\begin{smallmatrix}
	1 & 0 \\ 0 & e^{i\pi/4}
	\end{smallmatrix}\right)$, $X^{1/2}=\frac 12\left(\begin{smallmatrix}
	1+i & 1-i \\ 1-i & 1+i
	\end{smallmatrix}\right)$, and $Y^{1/2}=\frac 12\left(\begin{smallmatrix}
	1+i & -1-i \\ 1+i & 1+i
	\end{smallmatrix}\right)$.

The state of a two-qubit system $\Ket{\phi}$ can be represented using $4$ complex coefficients giving the contribution of all possible classical states featuring two bits, i.e.,
\[
	\Ket{\phi}=\alpha_{00}\Ket{00}+\alpha_{01}\Ket{01}+\alpha_{10}\Ket{10}+\alpha_{11}\Ket{11}=
	\left(\begin{matrix}
	\alpha_0\\\alpha_1\\\alpha_2\\\alpha_3
	\end{matrix}\right)
	\;.
\]
and operations acting on the entire state can be represented by complex unitary matrices of dimension $4\times 4$. An example for a two-qubit gate which occurs in quantum supremacy circuits is the controlled-Z or CZ gate,
\[
CZ=\left(\begin{matrix}
	1 & 0 & 0 & 0\\0 & 1 & 0 & 0\\0 & 0 & 1 & 0\\0 & 0 & 0 &-1
	\end{matrix}\right)\;,
\]
which adds a $(-1)$-phase to the $\Ket{11}$ basis state. Also, note that this gate is symmetric in terms of qubits -- it does not matter which qubit is the control / target qubit. This symmetry of the CZ gate can also be seen in Fig.~\ref{fig:supremacy_circuit}.

To build the $4\times 4$ matrix acting on the entire system when applying a single-qubit operation to just one qubit, one performs a Kronecker product with a $2\times 2$ identity matrix. Applying an $X$-gate to the first qubit (with bit-index 0) can be achieved as follows:
\begin{align*}
	X_0\Ket{\phi}&=\mathbbm 1_2\otimes X\Ket{\phi}\\
	&=\alpha_{01}\Ket{00}+\alpha_{00}\Ket{01}+\alpha_{11}\Ket{10}+\alpha_{10}\Ket{11}\\
	&=\alpha_{00}\Ket{01}+\alpha_{01}\Ket{00}+\alpha_{10}\Ket{11}+\alpha_{11}\Ket{10}\;.
\end{align*}
More generally, the state of an $n$-qubit quantum computer can be represented by a complex vector of size $2^n$
\[
	\Ket{\Psi} = \alpha_0\Ket{0\cdots 00} + \alpha_1\Ket{0\cdots01} +\cdots + \alpha_{2^n-1}\Ket{1\cdots 11}
\]
and operations on this state are $2^n\times 2^n$ unitary matrices. Finally, applying a single-qubit gate $U$ to the $i$-th qubit of an $n$-qubit quantum computer amounts to multiplying the state vector of coefficients $\alpha_i$ by the matrix
\[
	\underbrace{\mathbbm 1_2 \otimes\cdots\otimes \mathbbm 1_2}_{n-i-1} \otimes{}~ U \otimes \underbrace{\mathbbm 1_2 \otimes\cdots \otimes \mathbbm 1_2}_{i}\;,
\]
which is just a complex sparse matrix-vector multiplication of dimension $2^n$. Thus, for double-precision values, just storing the state vector for 50 qubits would already require 16 petabytes of memory.

\begin{figure*}[t]
\centering
\begin{subfigure}{.45\linewidth}
\includegraphics[width=\linewidth]{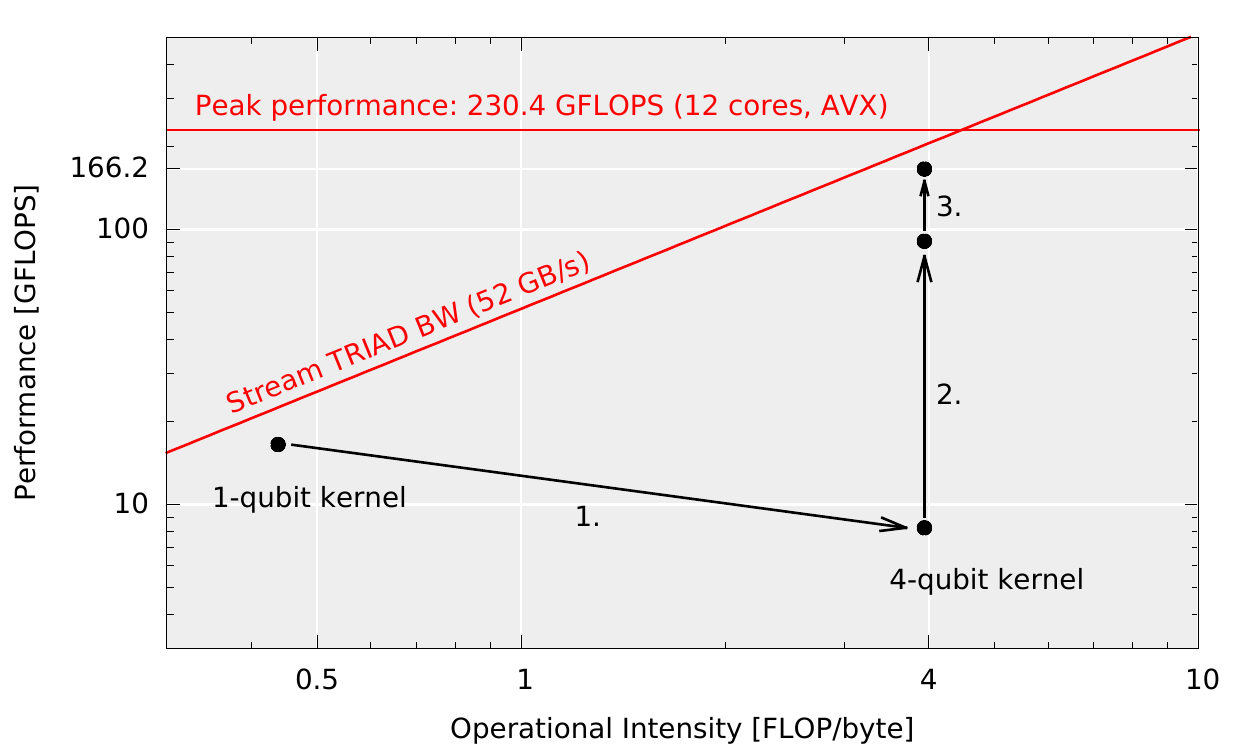}
\caption{Roofline plot for one Edison socket.}
\label{fig:rooflineedison}
\end{subfigure}
\begin{subfigure}{.45\linewidth}
\centering
\includegraphics[width=\linewidth]{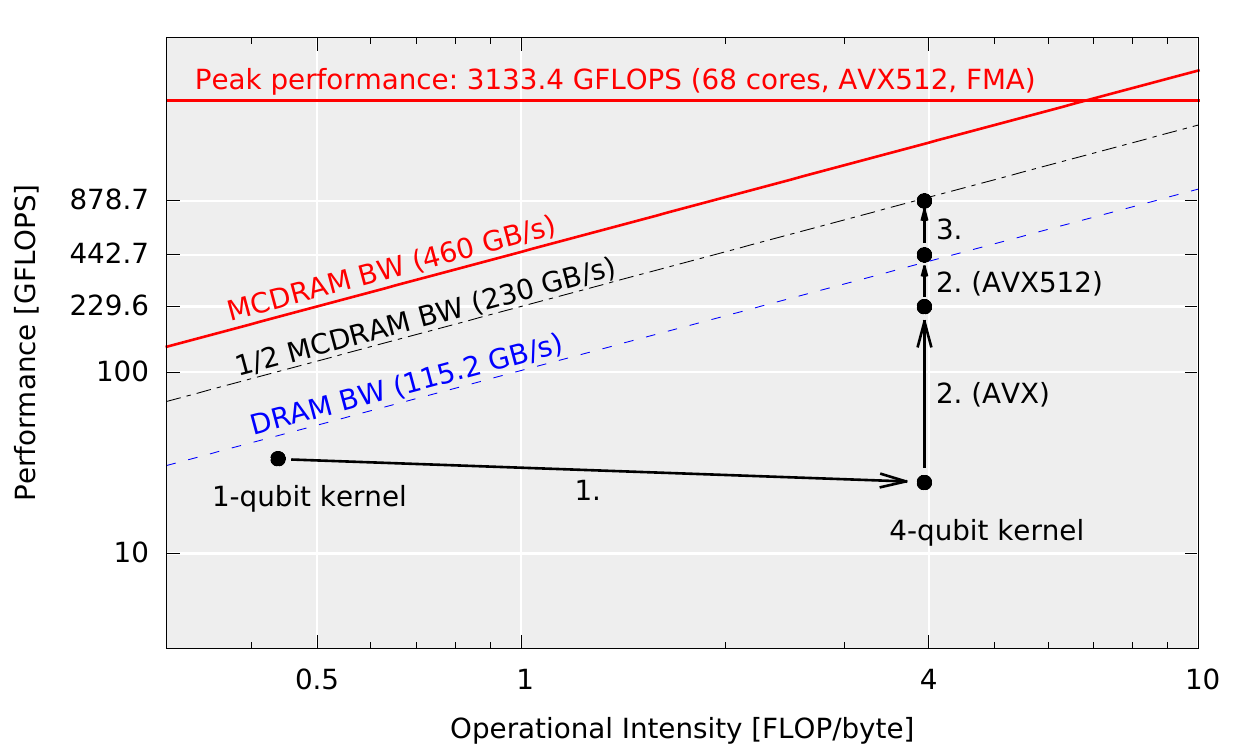}
\caption{Roofline plot for one KNL node of Cori II. }
\label{fig:rooflineknl}
\end{subfigure}
\caption{Roofline plots illustrating the performance improvements from Sec.~\ref{sec:singlecore} and~\ref{sec:singlenode}. Step~1 introduces lazy evaluation, making the application more compute-bound. Step~2 adds explicit vectorization and instruction re-ordering, followed by step~3 which applies blocking for registers in addition to a pre-computation on the gate matrix, re-ordering and permuting the complex-valued matrix entries to improve the FLOP/instruction ratio. An additional optimization specific to KNL is the blocking for MCDRAM, which is introduced in step 1.}
\end{figure*}

\section{Optimizations}
Our simulator was implemented and optimized using a layered approach. The first layer aims to improve the single-core performance of our quantum gate kernels by employing explicit vectorization using compiler intrinsics, instruction reordering, and blocking to reduce register-spilling. The second layer uses OpenMP to enable a good strong scaling behavior on an entire node. The third and final layer implements the inter-node communication scheme using MPI. This allows to simulate up to $45$ qubits on current supercomputers, in addition to reducing the time-to-solution when executing quantum circuits featuring fewer qubits.

\subsection{Standard optimizations}

In order to be able to simulate large systems, it is important not to actually store the $2^n\times 2^n$ matrix acting on the state vector. Instead, one can exploit its regular structure and implement methods which, given the state vector, mimic a multiplication by this matrix.
A standard implementation features two state vectors (one input, one output). To determine one entry of the output vector, two complex multiplications and one complex addition have to be carried out on two entries of the input vector when applying a general single-qubit gate. In total, there are thus
\[
	2\cdot (4\text{[mul]} + 2\text{[add]}) + 2\text{[add]} = 14\text{ FLOP}
\]
per complex entry of the output state vector. One complex double-precision entry requires 16 bytes of memory and the input vector has to be loaded from memory and the output vector has to be written back to memory. The operational intensity is therefore less than 1/2, which shows that this application is memory-bandwidth bound on most systems.

\subsection{Single-core}\label{sec:singlecore}
In order to reduce the memory requirements by a factor of 2x, this complex sparse matrix-vector multiplication can be performed in-place, at the cost of a cache-unfriendly access pattern. Moreover, $k$-qubit gates require more operations for larger $k$, allowing to better utilize hardware with strong compute capabilities. In fact, the number of operations grows exponentially with $k$, since applying a $k$-qubit gate amounts to performing one scalar product of dimension $2^k$ per (output) entry.

To apply a $k$-qubit gate (of dimension $2^k\times 2^k$) to a state vector of size $2^n$, where $n$ denotes the number of qubits, the entries corresponding to all $2^k$ indices of the gate matrix have to be loaded into a $2^k$-sized temporary vector, which then gets multiplied by the matrix before it is written back to the state vector. The indices of these state vector entries, when represented in binary, are bit-strings of the form $c_{n-k-1}x_{i_{k-1}}...c_{j}...x_{i_1}...c_{0}$, where $i_0,i_1,...,i_{k-1}$ denote the $k$ qubits indices to which the gate is being applied. Extracting and combining the bits $x_{i_j}$ from the index of an entry, i.e.,
\[
	x=x_{i_{k-1}}...x_{i_1}x_{i_0}\;,
\]
yields the index of this entry with respect to the temporary vector. All $2^k$ entries which have an identical $c=c_{n-k-1}c_{n-k-2}...c_{0}$ index substring are part of this matrix-vector multiplication. Once all entries have been gathered, multiplied by the matrix, and stored back into the state vector, the next $c'=c+1$ index substring can be dealt with. In total, this amounts to performing $2^{n-k}$ complex matrix-vector multiplications of dimension $2^k$.

A first observation is that the same matrix is used $2^{n-k}$ times. One can thus permute the matrix entries before-hand in order to always have sorted qubit indices, which results in memory accesses to occur in a more local fashion.

When applying the matrix-vector product, doing so in the usual manner, i.e.,
\[
	\tilde v_l = \sum_{i=0}^{2^k-1} m_{l,i} v_i\;,
\]
would require all entries of the temporary vector $v$ to be in register (and already loaded from memory). In order to address this issue, we employ blocking of the computation and determine the block size using an automatic code-generation / benchmarking feedback loop. For each block index $b=0,1,...,\frac{2^k}B-1$, all indices $l$ of the temporary output vector $\tilde v$ are updated according to
\[
	\tilde v_l \mathrel{+}= \sum_{j < B} m_{l,i(b,j)} v_{i(b,j)}\;,
\]
where $i(b,j)=b\cdot B+j$, before moving on to the next block.

We employ explicit vectorization to parallelize updates for consecutive values of $l$. Since we are dealing with complex double-precision values, this theoretically allows to speed up the execution by a factor of $2x$ or even $4x$ when using AVX or AVX512, respectively. Denoting by $a_R$ and $a_I$ the real and imaginary parts of $a$, respectively, we now inspect the update above more closely.
Multiplying one complex entry $v_l=(v_R, v_I)$ of the temporary vector $v$ with one complex entry of the gate matrix $m=(m_R, m_I)$ and summing the result into the temporary output vector $\tilde v$ can be written as follows:
\begin{align}
	(\tilde v_R, \tilde v_I) &\mathrel{+}= (v_R\cdot m_R - v_I\cdot m_I, v_I\cdot m_R+v_R\cdot m_I)
\end{align}
Yet, implementing this update results in wasted compute resources due to artificial dependencies and additional permutes. However, these instructions can be re-ordered as follows
\begin{align}
	(\tilde v_R, \tilde v_I) &\mathrel{+}= (v_R\cdot m_R, v_I\cdot m_R)\\
	(\tilde v_R, \tilde v_I) &\mathrel{+}= (v_I\cdot -1\cdot m_I, v_R\cdot m_I)
\end{align}
in order to increase the maximal achievable performance. Namely, having both $(m_R,m_R)$ and $(-1\cdot m_I, m_I)$ available, this update requires only two fused multiply-accumulate instructions instead of several individual multiplications, additions, and permutations. This is an improvement in both FLOP/instruction and FLOP/FMA ratios.

Note that $v_l$ can be permuted once upon loading (and then kept in register), as it is re-used for $2^k$ such complex multiplications. Also, since the matrix $m$ is used in $2^{n-k}$ matrix-vector multiplications, the pre-computation to build up these two matrices consisting of $(m_R,m_R)$ and $(-1\cdot m_I, m_I)$ is essentially free.

\subsection{Single-node}\label{sec:singlenode}

The optimizations discussed above do not change the fact that the operational intensity for applying a $1$-qubit gate is very low, making it harder to fully utilize the power of multi- and manycore processors (see, e.g., Fig.~\ref{fig:strongnodescalingedison}). Yet, as mentioned previously, applying a $k$-qubit gate requires more operations for larger values of $k$ and as long as the application remains memory bound, larger gates can be applied in (almost) the same amount of time. The benefit -- besides increased operational intensity -- is that larger gates can be used to execute an entire sequence of single- and two-qubit gates at once. In particular, multiple gates acting on $k$ different qubits can be combined into one large $k$-qubit gate. 

Which value of $k$ to choose depends on the peak performance, the memory-bandwidth, the cache-size \& associativity of the system, and the circuit to simulate. The cache specifications are important especially when gates are applied to qubits with larger indices, which cause memory access strides of large powers of two. For low-associativity caches, this causes conflicts to arise already for small kernel sizes. Since $2^k$ values need to be loaded from the state vector (which are at least $2^m$ apart, where $m$ is the lowest qubit index) for each of the $2^{n-k}$ matrix-vector multiplications, a $2^k$-way cache should map the corresponding cache-lines to different locations, no matter how large $m$ is. This allows to directly access these values from cache for the next matrix-vector multiplication. See Fig.~\ref{fig:hilow} and Fig.~\ref{fig:hilowedison} for experimental results.

Finally, these $k$-qubit gate kernels are parallelized using OpenMP with NUMA-aware initialization of the state vector to ensure scaling beyond 1 NUMA node. Depending on the qubits to which the gate is applied, the outer-most loop may perform very few iterations, prohibiting a good strong scaling behavior. The OpenMP \texttt{collapse} directive remedies this problem.

Please see Fig.~\ref{fig:rooflineedison} and Fig.~\ref{fig:rooflineknl}, which show the improvements in performance when applying all mentioned optimizations and running the kernels on one socket of Edison or Cori II, which feature one 12-core Intel\textregistered{} Xeon\textregistered{} Processor E5-2695 v2 and one 68-core Intel\textregistered{} Xeon Phi\texttrademark{} Processor 7250 (KNL), respectively.

\subsection{Multi-node}\label{sec:multinode}
The simulation of quantum computers featuring many more than 30 qubits requires multiple nodes in order for the state vector to fit into memory. We use MPI to communicate between $2^g$ nodes, each node having its own state vector of size $2^l$, where $g$ and $l$ denote the number of global and local qubits, respectively. Gate operations on local qubits, i.e., qubits with index $i<l$, require no communication. Qubits with index $i\geq l$, on the other hand, do require communication.

There are two basic schemes which can be used to perform multi-node quantum circuit simulations. The first \cite{trieu2009large} keeps global qubits global and applies global gates by employing 2 pair-wise exchanges of half the state vector. The second scheme \cite{de2007massively} swaps global qubits with local ones, applies gates to local qubits in the usual fashion and, if need be, swaps them again with global qubits.
Note that swapping in a global qubit and then immediately swapping it back out requires the same amount of communication as the first scheme. We thus expect the global-to-local scheme to perform better and focus on this scheme.

\begin{figure}[t]
\begin{center}
\begin{subfigure}{\linewidth}
\centering
\includegraphics[width=.43\linewidth]{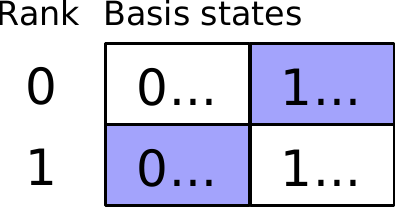}
\vspace{10pt}
\caption{Single-qubit swap.}
\label{fig:swap1}
\end{subfigure}

\vspace{10pt}

\begin{subfigure}{\linewidth}
\centering
\includegraphics[width=.75\linewidth]{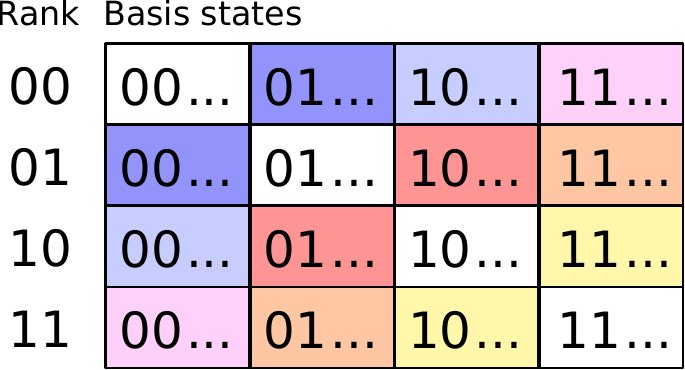}
\vspace{5pt}
\caption{Two-qubit swap.}
\label{fig:swap}
\end{subfigure}
\end{center}
\caption{Illustration of a single- and multi-qubit global-to-local swap using one (group-) all-to-all. The blocks labeled, e.g., $01...$ represent the coefficients corresponding to the global basis state which starts with the bit-string $r01$, where $r$ is the bit-representation of the rank (see text).}
\end{figure}

\paragraph{1-Qubit Example (see Fig.~\ref{fig:swap1}).} For the case of two ranks, swapping the highest-order qubit (highest bit in the local index) with the global qubit (first bit of the rank number) can be achieved as follows: The first block of rank 0 remains unchanged, since swapping 0 with 0 has no effect. Swapping 0 (global) and 1 (local) for the second block requires sending the entire block to rank 1, where these coefficients are associated with the local qubit being 0. Proceeding in this manner results in an exchange of the colored blocks, which is equivalent to an all-to-all.
\paragraph{2-Qubit Example (see Fig.~\ref{fig:swap}).} To swap two global qubits with the two highest-order local qubits for the case of four ranks, each rank sends its $i$-th quarter of the state vector to rank number $i$. Therefore, all identically-colored state vector parts are exchanged, which results again in one all-to-all.\\

Additionally, as done in~\cite{de2007massively}, we generalize this scheme to swap multiple or even all global qubits with local ones. Yet, in contrast to \cite{de2007massively}, we do not iteratively copy out parts of the state vector and carry out the pair-wise exchanges manually. Instead, we employ higher-level abstractions to achieve the same task, with the benefit that optimized implementations for, e.g., specific network topologies are likely to be already available. A $q$-qubit global-to-local swap, which exchanges $q$ global with $q$ local qubits, can be achieved using 1 group-local all-to-all for each of the $2^{g-q}$ groups of processes. Therefore, turning all global qubits into local ones amounts to executing one all-to-all on the \texttt{MPI\_COMM\_WORLD} communicator. This allows swapping the $k$ qubits with highest local index with $k$ global ones. In order to allow for arbitrary local qubits to be exchanged, we first use our optimized kernels to achieve local swaps between highest-index qubits and those to be swapped. We then perform the group-local all-to-all and, if need be, another local swap (with lower-index qubits) in order to improve data locality in our $k$-qubit gate kernels.

\subsection{Global gate specialization}
While a general global gate always requires communication, there are a few common ones which do not. Examples include the controlled-NOT gate (or controlled-X) which, when applied to global qubits, causes merely a re-numbering of ranks. The (diagonal) controlled-Z gate either turns into a conditional global phase or a local Z-gate which, depending on the rank, is executed or not. Finally, the T-gate is also diagonal and results in a global phase, which can be absorbed into the next gate matrix to be applied. Making use of such insights allows to further reduce the number of global-to-local swaps without increasing the amount of computation performed locally.

For 36-qubit quantum supremacy circuits, this optimization enables a reduction of the required communication by another factor of $2x$: Only one global-to-local swap is required to run the entire depth-25 circuit. For $42$- and $45$-qubit circuits, 2 global-to-local swaps are necessary, whereas 3 are required without gate specialization.

\subsection{Circuit Optimizations: Gate scheduling and qubit mapping}\label{sec:scheduling}

In addition to performing implementation optimizations, also the circuit requires optimization in order to reduce the number of communication steps and to use our highly-tuned kernels in a more efficient manner. We will demonstrate the different optimizations for gate scheduling and qubit mapping using the quantum supremacy circuits from~\cite{boixo2016characterizing}, for which we also present performance results in the next section. Our optimizations are general and can be applied to any quantum circuit. In fact, these quantum supremacy circuits happen to be designed in a way that is least suitable for these kinds of performance optimizations. We thus expect even larger improvements when employing these techniques for the simulation of other circuits.

The construction of these random, low-depth quantum circuits is shown in Fig.~\ref{fig:supremacy_circuit}. These circuits are designed to be run on a quantum computer architecture featuring a 2D nearest-neighbor connectivity graph. By design, all possible two qubit gates are applied within 8 cycles, which makes the system highly entangled. Note that a simulator can skip the initial Hadamard gates in cycle 0 and initialize the wave function directly to $(2^{-\frac n2},...)^T$, instead of starting in state $\Ket{0...0}=(1,0,...,0)^T$. Furthermore, we do not simulate the final CZ gates as they only alter the phases of the probability amplitudes $\alpha_i$, but not the probabilities $p_i=|\alpha_i|^2$ which we are interested in.

\subsubsection{Gate scheduling}
The most important optimization on the quantum circuit is gate scheduling, as it drastically reduces the amount of communication in the multi-node setting and also the number of $k$-qubit gate kernels on the single-node level. The optimizations can be broken into three steps:

\paragraph{1. Minimize number of communication steps}
In a first optimization step, gate scheduling minimizes the number of global-to-local swaps which is the most important parameter in the multi-node setting. Executing every clock-cycle of the circuit on its own requires at least one communication step for every cycle which features a non-diagonal global gate. 
 
\begin{figure}[ht]
	\includegraphics[width=\linewidth]{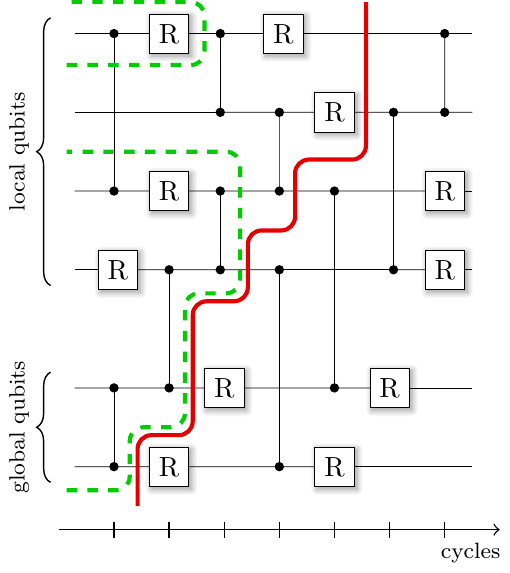}
	\caption{Example of gate scheduling for a circuit with CZ gates and dense single-qubit rotations gates (R). Note that we use gate specialization for CZ gates, which means we can apply them without communication on global qubits. First, instead of applying the gates cycle by cycle, we identify the largest first stage of gates which can be applied without communication. These are all the gates on the left of the solid red line. Second, we schedule the gates within a stage into clusters. For example, we can combine all the gates on the left of the dashed green line into one 3-qubit gate instead of applying 7 individual gates.}
	\label{fig:gate_scheduling}
\end{figure}

\begin{figure*}
	\begin{subfigure}{.45\linewidth}
	\centering
	\includegraphics[width=\linewidth]{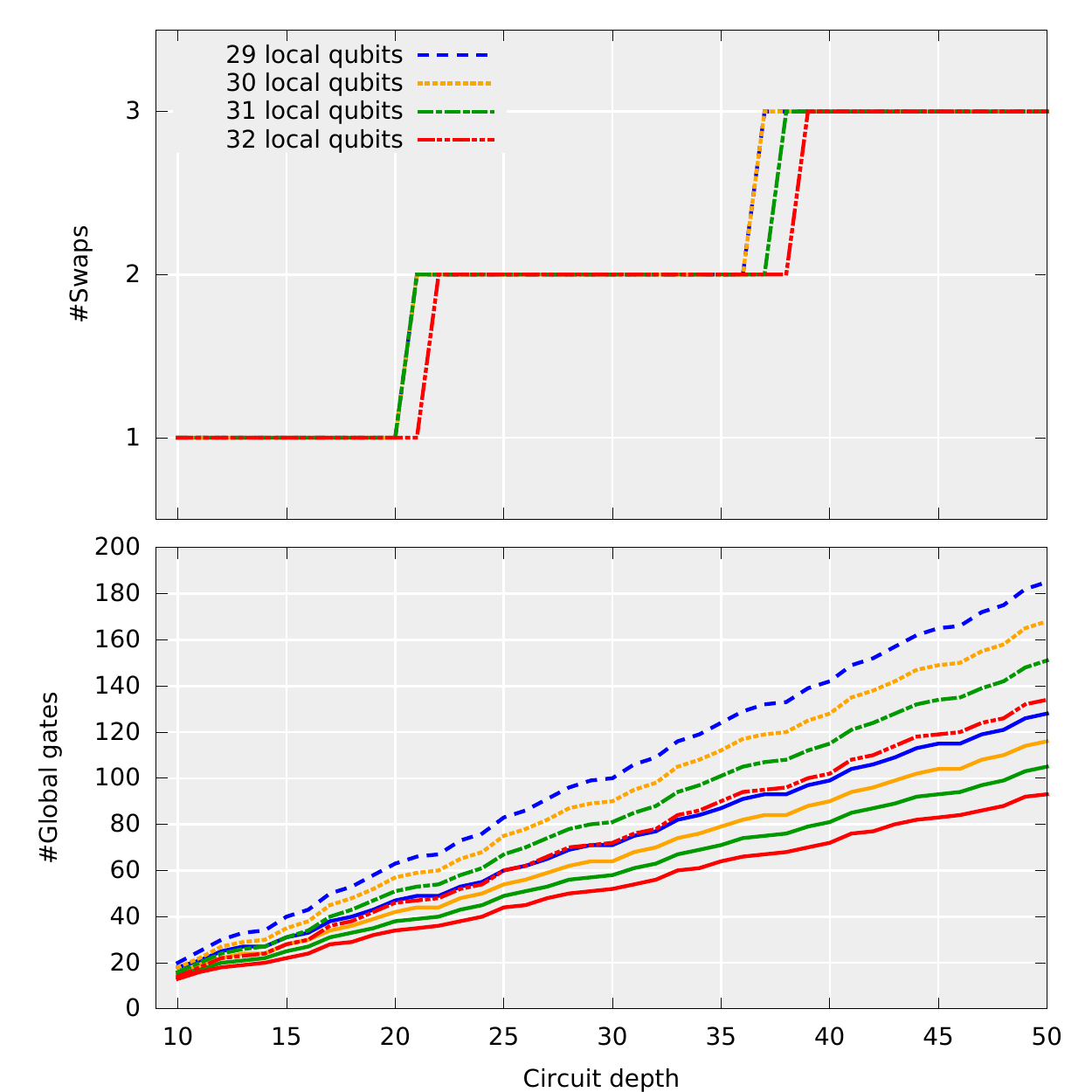}
	
	\caption{Scaling of the required communication for circuit depths $10$ to $50$ for $42$-qubit quantum supremacy circuits. The number of global-to-local swaps is mostly independent of the number of local qubits per node, which allows for a good strong scaling behavior.}
		          
	\label{fig:swaps_vs_global_gates}
	\end{subfigure}\hspace{20pt}
	\begin{subfigure}{.45\linewidth}
	\centering
	\includegraphics[width=\linewidth]{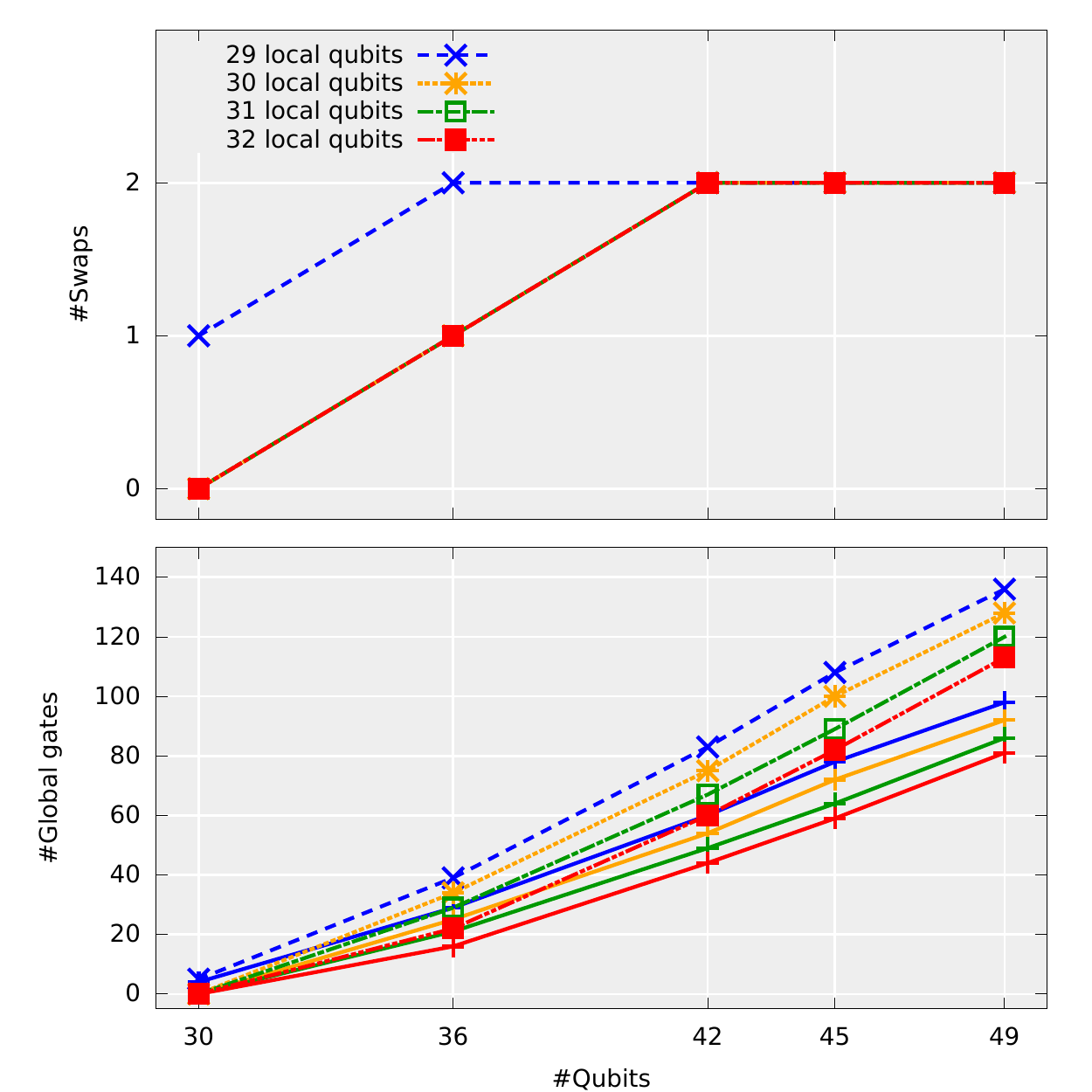}
	\caption{Scaling of the required communication for increasing numbers of qubits for quantum supremacy circuits with a fixed circuit depth of 25.
		\vspace{20pt}}
	
	\label{fig:swaps_scaling_with_problem_size}
	\end{subfigure}
	\caption{Scaling of the required number of communication steps for quantum supremacy circuits as a function of circuit depth (a) or number of qubits (b). The lower two panels show the number of global gates which require communication if executed individually as in \cite{boixo2016characterizing}. In contrast, the top two panels show the number of global-to-local swaps required to execute the full circuit when using our strategy of reordering gates and swapping global with local qubits. Note that one global-to-local swap (of all global qubits) requires the same amount of communication as one global gate. Averaged over the different global qubits, executing a dense global gate takes approximately $1/2$ of the time required to swap all global qubits with local qubits, because applying a dense gate to low-order global qubits is faster due to the increased locality of the communication, see \cite{boixo2016characterizing}. Note that the dashed lines are for worst case instances (only dense random gates on global qubits) and solid lines are for median hard instances, which we only consider in the two lower panels.}
\end{figure*}

However, as explained in the multi-node strategy, it is beneficial not to execute those global gates but rather swap global qubits with local qubits and then execute these gates locally. In order for this scheme to be most beneficial, the gate scheduling algorithm reorders (if possible) the gates into stages, where each stage consists of a sequence of quantum gates acting only on local qubits, see Fig.~\ref{fig:gate_scheduling}. Gates acting on the same qubit never commute for quantum supremacy circuits by design, making classical simulation harder. Nevertheless, we can reorder gates which act on different qubits as they commute trivially.
After completing a stage, some local qubits are swapped with global qubits, and a new stage is started. This scheme reduces the number of communication steps significantly. A depth-25 42-qubit supremacy circuit requires only two global-to-local swaps, see Fig.~\ref{fig:swaps_scaling_with_problem_size}. An important feature of our gate scheduling algorithm is that the number of global-to-local swaps is mostly independent of the number of local qubits (29, 30, 31, or 32). This allows for a good strong scaling behavior. Fig.~\ref{fig:swaps_vs_global_gates} shows how the number of global-to-local swaps behaves as a function of circuit depth.

We decided to always swap global qubits with the lowest-order local qubits to arrive at an upper bound for the number of communication steps required. In addition, we apply a cheap search algorithm to find better local qubits to swap with. In case of a 36-qubit supremacy circuit, this results in a $2x$ reduction in the number of global-to-local swaps, from two swaps to just one. Note that our stage-finding algorithm assumes the worst-case scenario, in which all randomly picked global single-qubit gates are dense, meaning that we cannot apply our gate specialization for T gates to reduce the amount of communication.

\paragraph{2. Minimize number of $k$-qubit gates}
In a second step, we schedule all the gates within a stage such that we can merge sequences of consecutive 1- or 2-qubit gates into a $k$-qubit gate and execute this $k$-qubit gate instead of many single- and two-qubit gates. See Fig.~\ref{fig:gate_scheduling}, which shows how such a cluster with $k=3$ can be built. We greedily try to increase the number of qubits $k$ within a cluster while still maintaining the condition that $k\leq k_{max}$, where $k_{max}$ is the largest $k$ for which the $k$-qubit gate kernel still shows good performance on the target system. To reduce the over-all number of clusters, we perform a small local search in order to build the largest cluster with gates not yet assigned, before assigning the remaining gates to new clusters. We summarize the required number of clusters to execute a quantum supremacy circuit in Table~\ref{tab:clustering}. Clearly, even for these circuits, more than $k$ gates can be merged into one $k$-qubit cluster on average.

\paragraph{3. Local adjustments of global-to-local swaps}
The last cluster within each stage tends to contain a lower number of single- and two-qubit gates. In order to increase the average number of gates in each cluster and thereby decrease the total number of clusters in the circuit, we try to remove the last clusters of each stage by performing the global-to-local swap earlier if this is possible without increasing the total number of global-to-local swaps. 

\subsubsection{Qubit mapping}
Last, the bit-location of each qubit is optimized in order to reduce the number of clusters experiencing the performance decrease resulting from the set-associativity of the last-level cache. Since this performance decrease only occurs if the gate is applied to high-order bit-locations, this can be achieved by remapping. The following heuristic allowed for a $2x$ decrease in time-to-solution:

Assign the qubit to bit-location 0 such that the number of clusters accessing bit-location $0$ is maximal. From now on, ignore all clusters which act on this qubit and assign bit-locations $1$, $2$, and $3$ in the same manner. Bit locations $4$, $5$, $6$, and $7$ are assigned the same way, except that after each step, only clusters acting on two of these four bit-locations are ignored when assigning the next higher bit-location.
For non-random circuits, it would pay off to perform a few local swaps between some bit-locations over the course of the algorithm, in order to maximize the number of clusters acting on low-order qubits.

\renewcommand{\arraystretch}{1.6}
\begin{table}[ht]
	\centering
	\begin{tabular}{c@{\;\;\;\;\;\;}c@{\;\;\;\;\;\;}c@{\;\;\;\;\;}c@{\;\;\;\;\;}c@{\;\;\;\;\;}}\hlinewd{1pt}
		\multirow{2}{*}{\pbox{3cm}{Number\\\centering of Qubits}} & \multirow{2}{*}{\pbox{3cm}{Number\\\centering of Gates}} &\multicolumn{3}{c}{Number of clusters}\\ \cmidrule{3-5}
		&  & $k_{max}=3$ & $k_{max}=4$& $k_{max}=5$\\ \hlinewd{.5pt}

		30 & 369 & 82 & 46 & 36 \\
		36 & 447 & 98 & 53 & 41  \\
		42 & 528 & 111 & 58 & 46 \\
		45 & 569 & 111 & 73 & 51 \\ \hlinewd{1pt}\addlinespace[\belowrulesep]
		
	\end{tabular}
	\caption{Re-scheduling of gates for depth-25 quantum supremacy circuits into clusters (using 30 local qubits). Clusters are built to contain $k\leq k_{max}$ qubits using a heuristic which tries to maximize the number of gates merged into one cluster. Clearly more than $k_{max}$ individual gates can be combined into one single cluster on average. These optimizations take less than 3 seconds using Python and can be reused for all instance of the same size.}
	\label{tab:clustering}
\end{table}

\section{Implementation and Results}
All optimizations mentioned in the previous sections were implemented in C++, except for the code generator for the $k$-qubit kernels and the circuit scheduler/qubit mapper, which were both implemented in Python.

\subsection{Cori II}
We performed simulations of quantum supremacy circuits featuring $30$, $36$, $42$, and $45$ qubits on the Cori~II system at the Lawrence Berkeley National Laboratory (LBNL). {Cori~II} consists of 9,304 single-socket compute nodes, each containing one 68-core Intel\textregistered{} Xeon Phi\texttrademark{} Processor 7250 (KNL) at 1.40GHz. The nodes are interconnected by a Cray Aries high speed "dragonfly"~\cite{kim2008dragonfly} topology interconnect and offer a combined theoretical peak performance of 29.1 PFLOPS and 1 PB of aggregate memory.

\subsubsection{Node-level performance}

\begin{figure}[t]
	\includegraphics[width=\linewidth]{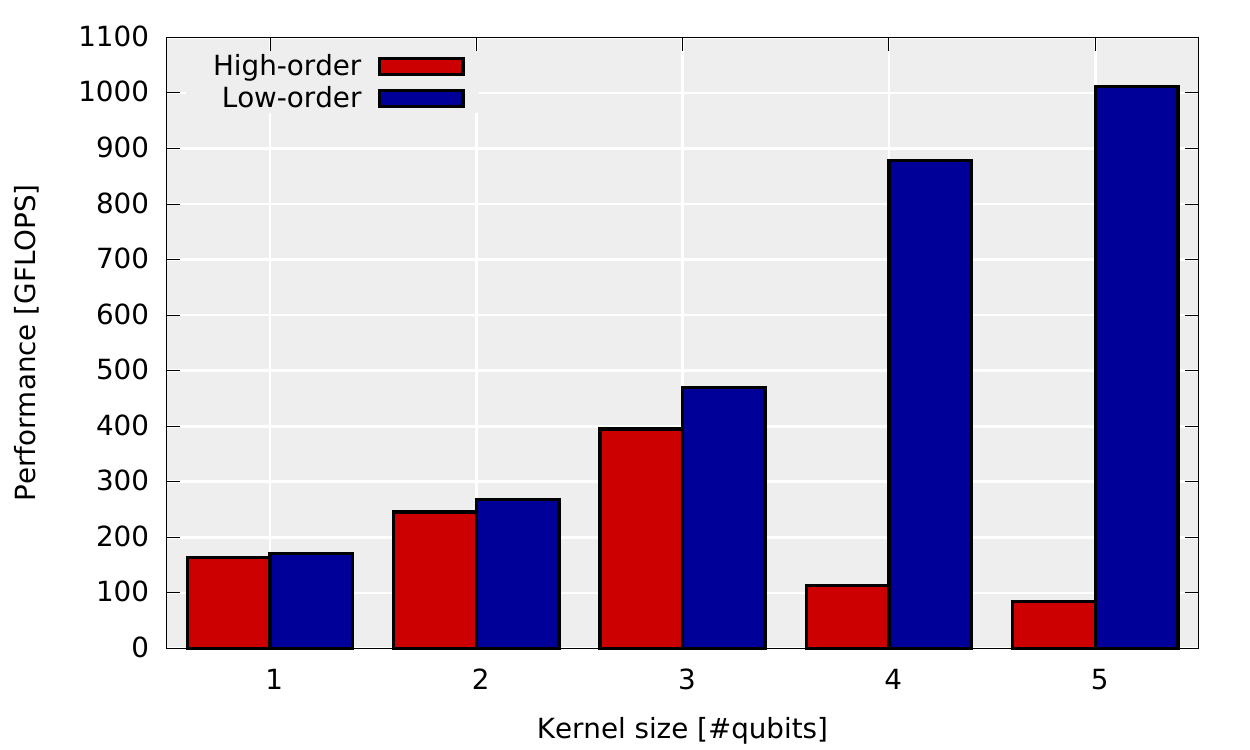}
	\caption{Decrease in performance when applying $k$-qubit gate kernels to qubits with large indices (high-order qubits) as opposed to low indices (low-order qubits). These experiments were run on all 68 cores of a Cori II KNL node. As mentioned in Sec.~\ref{sec:singlenode}, this performance drop occurs when $2^k$ is larger than the set-associativity of the last-level cache. While the L2-cache is $16$-way set-associative, it is shared between 2 cores.}
	\label{fig:hilow}
\end{figure}

\begin{figure}[t]
	\includegraphics[width=\linewidth]{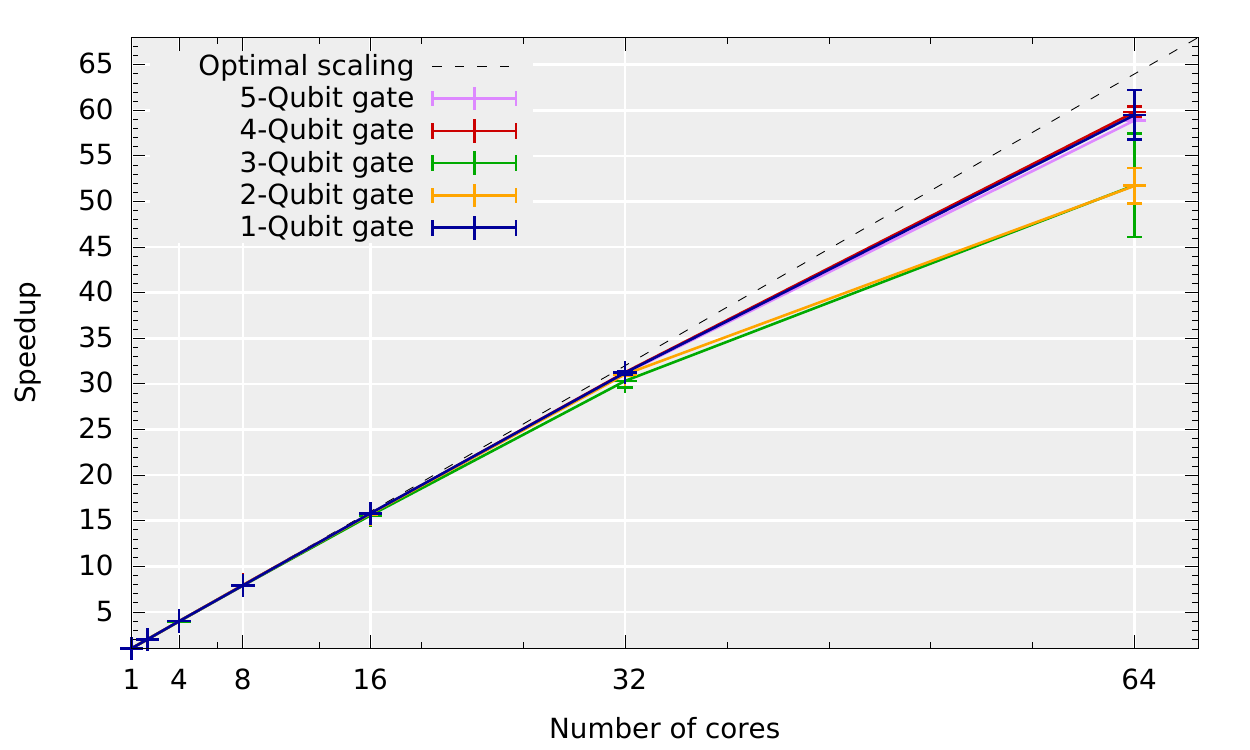}
	\caption{Strong scaling for applying $k$-qubit kernels to a 28-qubit system using $2^p$, $p\in\{0,...,6\}$ cores of the 68-core Intel\textregistered{} Xeon Phi\texttrademark{} Processor 7250 and 4, 2, and 1 OpenMP thread(s) per core for $k\leq 3$, $k=4$, and $k=5$, respectively.}
	\label{fig:strongnodescalingknl}
\end{figure}

These experiments were run on a single 68-core Intel\textregistered{} Xeon Phi\texttrademark{} Processor 7250 (KNL) node of the Cori II supercomputing system in the quad/cache setting. For $k\in\{1,2,3\}$-qubit gate kernels, four threads per core were used, as this resulted in the best performance. For $k=4$ and $k=5$, the best performance was achieved when using two and one thread per core, respectively.
As mentioned in Sec.~\ref{sec:singlenode}, the set-associativity of caches plays a crucial role in the performance of these $k$-qubit gate kernels. In particular, we find the theoretical predictions from Sec.~\ref{sec:singlenode} to agree perfectly with observations, see Fig.~\ref{fig:hilow}.
The strong scaling behavior of executing one $k$-qubit gate kernel on a state vector of $28$ qubits can be seen in Fig.~\ref{fig:strongnodescalingknl}.

\subsubsection{Multi-node performance}

\begin{figure}[t]
	\includegraphics[width=\linewidth]{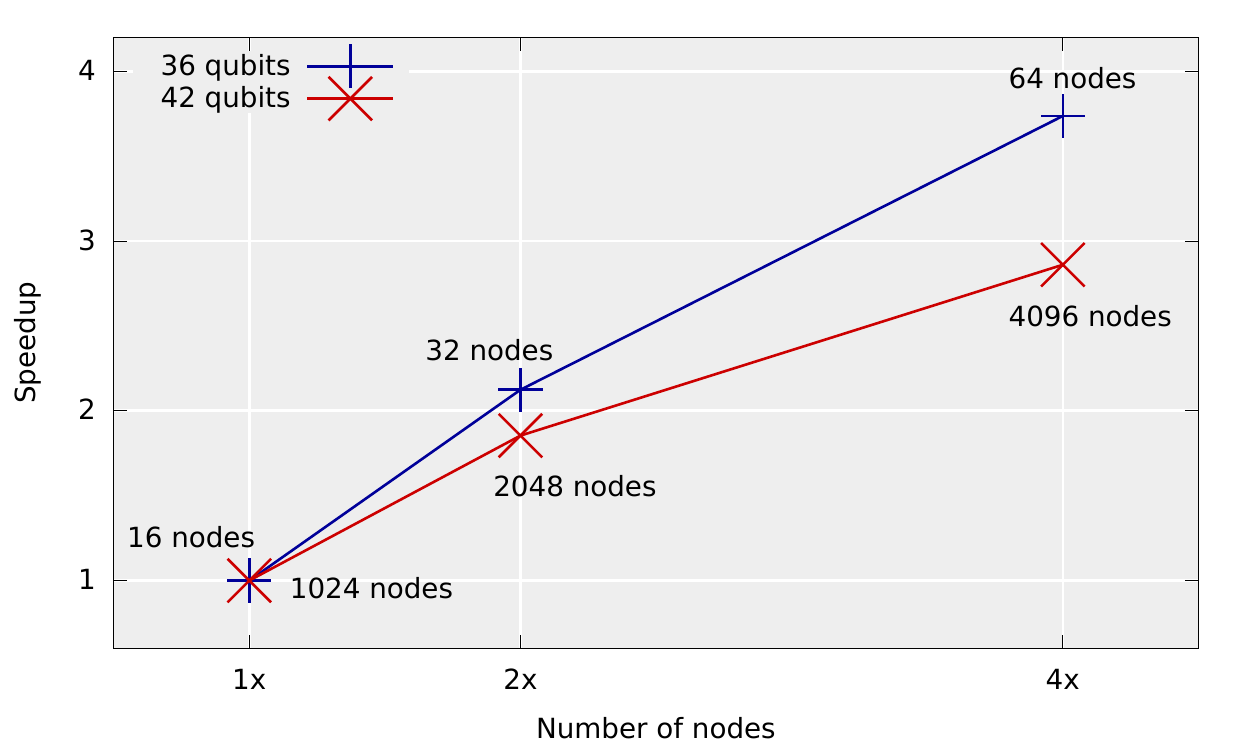}
	\caption{Strong scaling of our simulator running a 36- and 42-qubit quantum supremacy circuit on $\{16,32,64\}$ and $\{1024,2048,4096\}$ nodes of Cori II, respectively.}
	\label{fig:multinodescalingknl}
\end{figure}

The strong scaling of our simulator for a 36- and 42-qubit quantum circuit running on $\{16,32,64\}$ and $\{1024,2048,4096\}$ KNL nodes of Cori II, respectively, is depicted in Fig.~\ref{fig:multinodescalingknl}.
Following these scaling experiments, we ran a $45$-qubit quantum supremacy circuit using $8,192$ KNL nodes and a total of $0.5$PB of memory. To our knowledge, this is the largest quantum circuit simulation ever carried out. Averaged over the entire simulation time (i.e., including communication time), this simulation achieved $0.428$ PFLOPS. There are two reasons for this drop in performance. First, the time spent in communication and synchronization is $78\%$, and overlaying computation and communication would not improve this behavior due to the low $k$-qubit gate times (less than 1 second).

\renewcommand{\arraystretch}{1.6}
\begin{table}[t]
\centering
\begin{tabular}{lc@{\;\;\;\;}r@{\;\;\;\;\;}c@{\;\;\;\;}cc@{\;\;\;}}
\hlinewd{1pt}
\#Qubits& $\#$Gates & $\#$Nodes & Time [s] & Comm. & Speedup\\
\hlinewd{.5pt}
$6\times 5$ & $369$ & $1$ & $9.58$ & 0\% & $14.8x$\\
$6\times 6$ & $447$ & $64$ & $28.92$ & 42.9\% & $12.8x$\\
$7\times 6$ & $528$ & $4096$ & $79.53$ & 71.8\% &$12.4x$\\
$9\times 5$ & $569$ & $8192$ & $552.61$ & 78.0\% &$N/A$\\
\hlinewd{1pt}\addlinespace[\belowrulesep]
\end{tabular}
\caption{Results for all simulations carried out on Cori II. Circuit simulation time and speedup are given with respect to the depth-25 quantum supremacy circuit simulations performed in~\cite{boixo2016characterizing}. The comm.-column gives the percentage of circuit simulation time spent in communication and synchronization.}
\label{tbl:results}
\end{table}
Second, the performance of our kernels suffers in the regime where only few $k$-qubit gates are applied before a global-to-local swap needs to be performed. This is due to the fact that blocking for MCDRAM requires a sequence of several gates acting on qubits below bit-location 29. While our mapping procedure aims to maximize this number, the total number of gates being applied is not large enough. Yet, this is mainly due to the artificial construction of random circuits and does not occur in actual quantum algorithms, where interactions remain local over longer periods of time. As our 4-qubit gate kernel achieves $1/2$ of the MCDRAM bandwidth which corresponds to roughly $2x$ the bandwidth of DRAM (see Fig.~\ref{fig:rooflineknl}), we expect a $2x$ drop in performance if memory requirements exceed the MCDRAM size of 16GB. Averaging the performance of our $k$-qubit kernels in Fig.~\ref{fig:hilow} and including this $2x$ reduction yields approximately $250$ GFLOPS per node. In total, we thus expect a performance of $22\%\times8,192\times 250 \text{ GFLOPS}\approx 0.45 \text{ PFLOPS}$, which agrees with the measurement results given that we also apply a few $3$- and $2$-qubit gate kernels for left-over gates.

For a summary of all runs carried out on Cori II, see Table~\ref{tbl:results}. Our implementation for, e.g., 42 qubits behaves as expected from Fig.~\ref{fig:swaps_vs_global_gates}: For a depth-25 circuit, the communication scheme used in~\cite{boixo2016characterizing} requires about 50 global gates, while our simulator performs 2 global-to-local swaps (of all global qubits). Including the fact that one such global-to-local swap requires the same amount of communication and that, averaged over all global qubits, a global gate is $2x$ faster than if it is applied to the highest-order global qubit due to the network bisection bandwidth (see~\cite{boixo2016characterizing}), yields a reduction in communication of
\[
	\frac{50x}{2\cdot 2}=12.5x\;,
\]
and since we achieve a similar reduction in time-to-solution for the circuit simulation on each node, this is also the expected overall speedup.

\subsection{Edison}
In order to be able to compare our results directly to~\cite{boixo2016characterizing}, we also ran $30$- and $36$-qubit quantum supremacy circuits on the Edison system, also at LBNL. We used up to $64$ sockets, each featuring a $12$-core Intel\textregistered{} Xeon\textregistered{} Processor E5-2695 v2 at $2.4$GHz. The $5,586$ $2$-socket Edison nodes are interconnected by a Cray Aries "dragonfly"~\cite{kim2008dragonfly} topology interconnect and the theoretical peak performance of the entire system is $2.57$ PFLOPS.

\subsubsection{Node-level performance}
\begin{figure}[ht]
	\includegraphics[width=\linewidth]{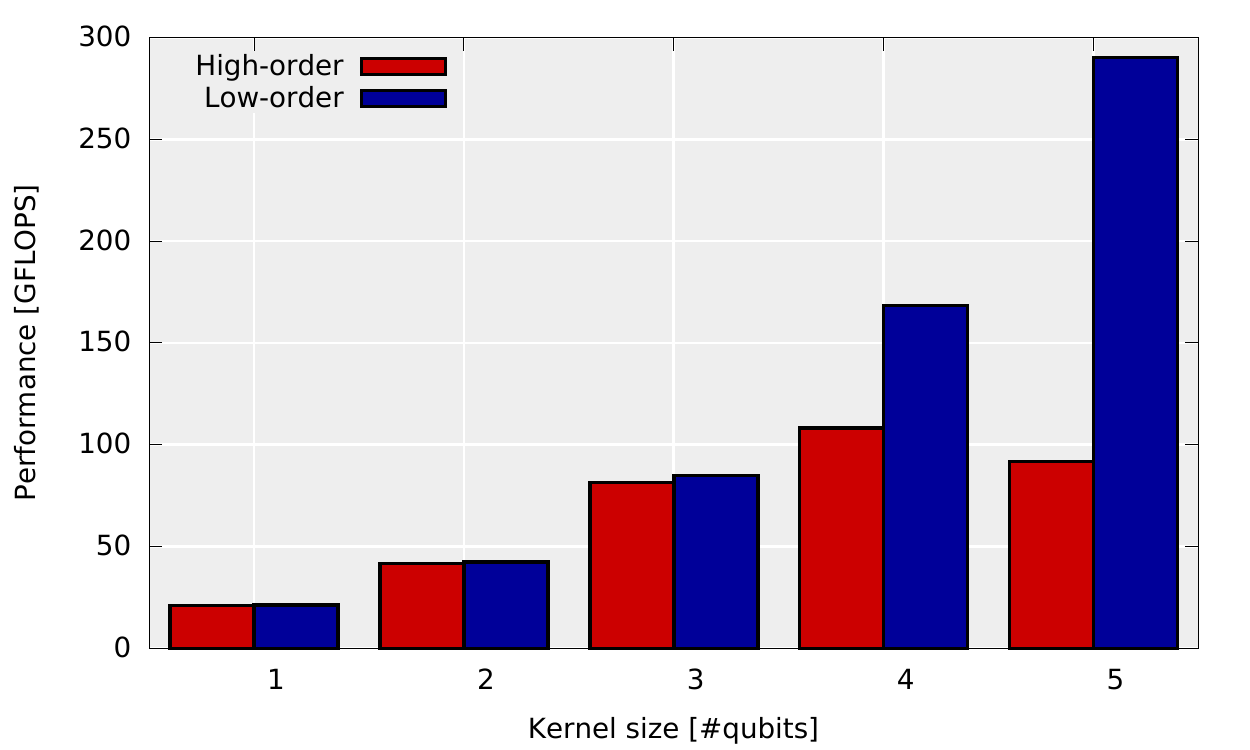}
	\caption{Performance decrease when $k$-qubit gate kernels are applied to high-order qubits instead of low-order ones on a two-socket Edison node. The findings again correspond to the set-associativity of the caches, which is $2^3=8$ in this case. For $k\leq 3$, there is only a negligible drop in performance, since all $2^k$ entries are mapped to different locations in the cache and the next $2^k$-sizes matrix-vector multiplication can access the next $2^k$ values directly from cache, see Sec.~\ref{sec:singlenode}}
	\label{fig:hilowedison}
\end{figure}

The performance reduction from applying gates to high-order qubits due to the $8$-way set-associativity of the L1- and L2-caches in Intel\textregistered{} Ivy Bridge\texttrademark{} processors can be seen in Fig.~\ref{fig:hilowedison}. These experiments were run on an entire two-socket node on all $24$ cores with one OpenMP thread per core and using AVX vectorization.

\begin{figure}[t]
	\includegraphics[width=\linewidth]{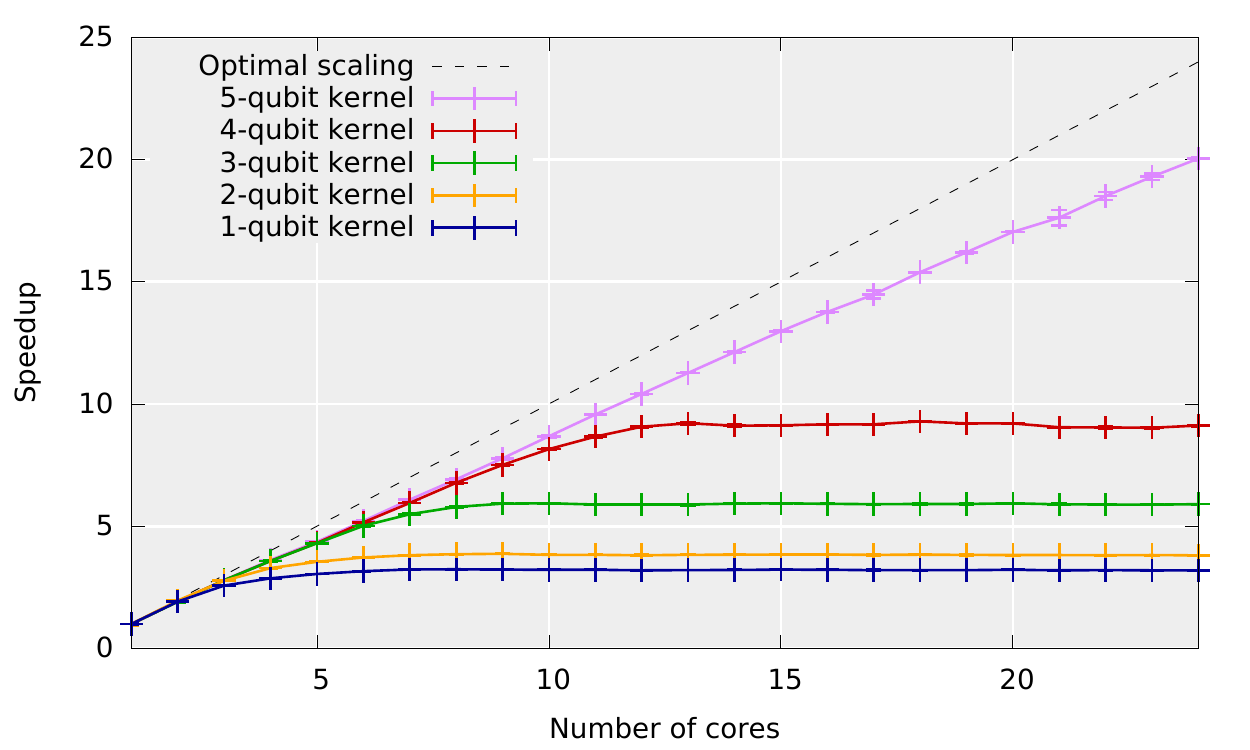}
	\caption{Strong scaling of the $k$-qubit kernels using up to 24 cores of a two-socket Edison node, which features one 12-core Intel\textregistered{} Ivy Bridge\texttrademark{} processor per socket. Up to and including $k=4$, the kernels are memory bandwidth limited. This in combination with Fig.~\ref{fig:hilowedison} suggests that $k=4$ is the best kernel size to use on this system (with 1 MPI process per socket).}
	\label{fig:strongnodescalingedison}
\end{figure}

The strong scaling of these $k$-qubit kernels with respect to the number of cores is depicted in Fig.~\ref{fig:strongnodescalingedison}. While the $5$-qubit gate kernel scales best to the full node, the performance drop when applying it to high-order qubits is much greater than it is for $4$-qubit gates. In addition, the $4$-qubit gate kernel scales nearly perfectly to all $12$ cores of a single socket, which suggests to use $2$ MPI processes per node in the multi-node setting.

Running a single-socket simulation of a $30$-qubit quantum supre\-macy circuit yields an improvement in time-to-solution by $3x$.

\subsubsection{Multi-node performance}

In order to compare the present work directly to the state-of-the-art simulator in~\cite{boixo2016characterizing}, we performed a simulation of a 36-qubit quantum supremacy circuit using identical hardware: 64 sockets of the Edison supercomputer. We calculated the entropy of a depth-25 quantum supremacy circuit in 99 seconds, of which 90.9 seconds were spent in actual simulation and the remaining 8.1 seconds were used to calculate the entropy, which requires a final reduction. This constitutes an improvement in time-to-solution of over $4x$ and indicates that the obtained speedups were not merely a consequence of a new generation of hardware.

The kernels perform at an average of $47\%$ theoretical peak, or $218$ GFLOPS on every node during the execution of a $36$-qubit quantum supremacy circuit. When including communication time, the entire simulation achieved $30\%$ of the theoretical peak performance of $64$ Edison sockets, which is $4.4$ TFLOPS.

\section{Summary and outlook}
We demonstrated simulations of up to 45 qubits using up to $8,192$ nodes. With the same amount of compute resources, the simulation of 46 qubits is feasible when using single-precision floating point numbers to represent the complex amplitudes. The presented optimizations are general and our code generator improves performance portability across a wide range of processors. Extending the range of the code generator to the domain of GPUs is an ongoing project.

Additional optimizations on the quantum circuit description allows to reduce the required communication by an order of magnitude. As a result, the simulation of a 49-qubit quantum supremacy circuit would require only two global-to-local swap operations. While the memory requirements to simulate such a large circuit are beyond what is possible today, the low amount of communication may allow the use of, e.g., solid-state drives. The simulation results may then be used for verification and calibration of near-term quantum devices.

\begin{acks}
We thank Jarrod McClean for his outstanding help and support throughout the course of this project. Special thanks goes to our advisor, Matthias Troyer, for giving us the opportunity to work on this project and for enlightening discussions. Moreover, we would like to thank Ryan Babbush, Sergio Boixo, Brandon Cook, Jack Deslippe, Sergei Isakov, and Hartmut Neven for helpful comments and discussions.

This research used resources of the National Energy Research Scientific Computing Center, a DOE Office of Science User Facility supported by the Office of Science of the U.S. Department of Energy under Contract No. DE-AC02-05CH11231. This work was supported by the Swiss National Science Foundation through the National Competence Center for Research NCCR QSIT.
\end{acks}

\bibliographystyle{ACM-Reference-Format}
\bibliography{references} 

\end{document}